# The regulatory switch of $F_1$-ATPase studied by single-molecule FRET in the ABEL Trap

Samuel D. Bockenhauer[a,b], Thomas M. Duncan[c], W. E. Moerner[a] and Michael Börsch[d,]*

[a] Department of Chemistry, Stanford University, Stanford, CA, USA
[b] Department of Physics, Stanford University, Stanford, CA, USA
[c] Department of Biochemistry & Molecular Biology, SUNY Upstate Medical University, Syracuse, NY, USA
[d] Single-Molecule Microscopy Group, Jena University Hospital, Friedrich Schiller University, Jena, Germany

## ABSTRACT

$F_1$-ATPase is the soluble portion of the membrane-embedded enzyme $F_oF_1$-ATP synthase that catalyzes the production of adenosine triphosphate in eukaryotic and eubacterial cells. In reverse, the $F_1$ part can also hydrolyze ATP quickly at three catalytic binding sites. Therefore, catalysis of 'non-productive' ATP hydrolysis by $F_1$ (or $F_oF_1$) must be minimized in the cell. In bacteria, the ε subunit is thought to control and block ATP hydrolysis by mechanically inserting its C-terminus into the rotary motor region of $F_1$. We investigate this proposed mechanism by labeling $F_1$ specifically with two fluorophores to monitor the C-terminus of the ε subunit by Förster resonance energy transfer. Single $F_1$ molecules are trapped in solution by an Anti-Brownian electrokinetic trap which keeps the FRET-labeled $F_1$ in place for extended observation times of several hundreds of milliseconds, limited by photobleaching. FRET changes in single $F_1$ and FRET histograms for different biochemical conditions are compared to evaluate the proposed regulatory mechanism.

## 1 INTRODUCTION

$F_oF_1$-ATP synthase is a multi subunit membrane enzyme that utilizes the electrochemical potential of protons (or $Na^+$ in some organisms) over the membrane to synthesize adenosine triphosphate (ATP) from adenosine diphosphate (ADP) and inorganic phosphate ($P_i$). Depending on physiological conditions, the bacterial enzyme can also work in reverse and can hydrolyze ATP to pump protons across the membrane[1]. Proton translocation in the $F_o$ portion of the enzyme and chemical synthesis or hydrolysis of ATP in the $F_1$ portion are coupled *via* mechanical rotation of subunits within $F_oF_1$-ATP synthase. This mechanism was first proposed by P. Boyer about 30 years ago (reviewed in [2]) and subsequently demonstrated by a variety of biochemical[3-7] and spectroscopic[8] methods as well as single-molecule imaging[9-17] and single-molecule FRET experiments[18-27].

The simplest form of the soluble $F_1$ portion (often called $F_1$-ATPase) is found in bacteria and consists of five different subunits with stoichiometry $\alpha_3\beta_3\gamma\epsilon$. The crystal structure of the *Escherichia coli* $F_1$ was recently resolved at a resolution of 3.26 Å (Fig. 1A)[28, 29]. The pseudohexagonal arrangement of three pairs of subunits α and β, i.e. $\alpha_3\beta_3$, forms the main body of $F_1$. Each subunit β provides an active nucleotide binding site, while the corresponding nucleotide binding sites on the α subunits are catalytically inactive. Subunits $\alpha_3\beta_3$ (together with subunit δ at the top of $F_1$, not shown in Fig. 1A) comprise a non-rotating stator complex[30]. Subunits γ and ε form the central stalk that can rotate within $\alpha_3\beta_3$ and connects to the membrane-embedded ring of 10 *c*-subunits of the $F_o$ portion. The *a* subunit of $F_o$ provides two proton half-channels. The two *b* subunits of $F_o$ connect the membrane part as a peripheral stator stalk to the top of $F_1$. Thus, the holoenzyme $ab_2c_{10}\alpha_3\beta_3\gamma\delta\epsilon$ can transfer the energy of the transmembrane electrochemical potential of protons *via* rotational movements of $c_{10}$-ε-γ to the nucleotide binding sites in $\alpha_3\beta_3$, where it is transformed into chemical bond energy of ATP.

................................................................

* Email: michael.boersch@med.uni-jena.de; http://www.m-boersch.org

During the biochemical assembly process of $F_oF_1$-ATP synthase in *E. coli* proton leakage through a membrane-embedded $F_o$ subcomplex without attached $F_1$ or the waste of ATP by fast-hydrolyzing $F_1$ portions in the cytosol has to be prevented. To control ATP hydrolysis *in vivo* the bacterial $F_oF_1$-ATP synthase is thought to be regulated by subunit ε, a 15 kDa subunit of the $F_1$ rotor[31]. Its N-terminal β-sandwich domain (NTD) binds to the lower portion of the coiled-coil domain of subunit γ and to the *c*-ring. The C-terminal domain (CTD) of ε comprises two α-helices[32, 33] that adopt an 'extended'-configuration in the recent *E. coli* $F_1$ structure (Fig. 1 A, B). This is an intrinsic inhibited conformation of $F_1$ with the γ-ε rotor stalled at a fixed angle. However, in the active enzyme, ε's CTD is thought to form a hairpin-folded state[34] with the C-terminal helices in a 'down'-configuration (Fig. 1C). To reactivate the enzyme from the ε-inhibited state, a higher activation energy is needed than from another inhibited state, the so-called MgADP-inhibited state[35]. An *E. coli* $F_oF_1$-ATP synthase with a deleted CTD of subunit ε showed not only a higher ATP hydrolysis (ATPase) activity compared to the wild type, but also a higher ATP synthesis activity[36]. Therefore, a large movement of the C-terminal helices of ε is thought to be a mechanical switch which controls the enzymatic activities of both $F_1$-ATPase as well as $F_oF_1$-ATP synthase.

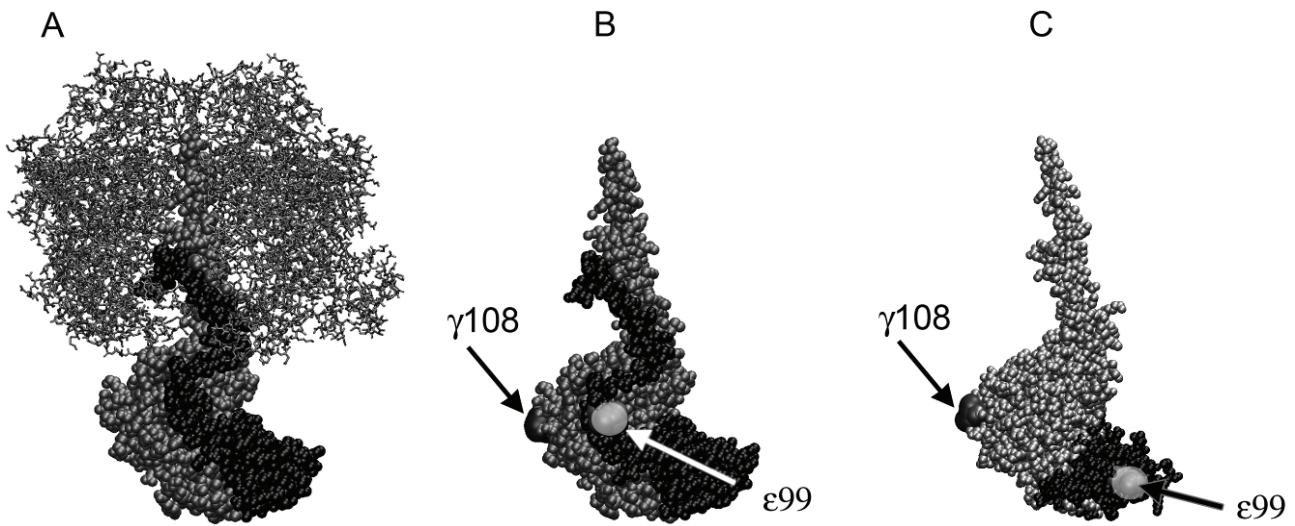

**Figure 1.** (**A**) Structure of *E. coli* $F_1$ with subunits $α_3β_3$ in grey (stick representation), γ in silver (ball representation) and ε in black (ball representation) in the 'up' configuration[28]. (**B**) Subunits γ and ε of *E. coli* $F_1$ as in (A), with positions of the two cysteine residues γ108 and ε99 for single-molecule FRET. (**C**) Partial structure of the γ- (in silver) and ε-subunits (in black) from $F_1$ with ε's C-terminal helices in the 'down'-configuration[33]. The residue positions of the two cysteines for single-molecule FRET between γ108 and ε99 are indicated.

To monitor conformational changes of ε's CTD in purified *E. coli* $F_1$-ATPase, we developed a single-molecule FRET approach. In this approach, two specifically attached fluorophores are used for an internal distance ruler based on FRET. Based on the *E. coli* $F_1$ X-ray structure[28], we chose residue ε99 on the first C-terminal α-helix of ε, which does not insert into a β-γ cleft in the 'up'-conformation (Fig. 1A). The second marker position is γ108, yielding distances of about 3 nm or 6 nm to ε99 in the 'up' (Fig. 1 B) or 'down' (Fig. 1C) conformations, respectively. These labeling positions were also chosen to avoid perturbing any interactions of ε CTD (either conformation) with the ε NTD or with other subunits.

However, a major limitation of a confocal single-molecule FRET approach using freely diffusing $F_1$-ATPase is the short observation time of a few milliseconds for $F_1$. To hold single small particles in solution the <u>A</u>nti-<u>B</u>rownian <u>E</u>lectrokinetic <u>T</u>rap (ABEL Trap) was invented by A. E. Cohen and W. E. Moerner[37]. The fast ABEL Trap captures liposomes, DNA, proteins or even single fluorophores in solution[37-41] using very fast feedback times in the microsecond time range[42-49].

Here we show that it is possible to hold a single FRET-labeled $F_1$ in solution for several hundred milliseconds and to measure the dynamics of the C-terminal domain of subunit ε.

## 2 EXPERIMENTAL PROCEDURES

### 2.1 Preparation of $F_1$-ATPase and the ε subunit for single-molecule FRET measurements

The ATP synthase from *Escherichia coli* was overexpressed with an N-terminal 6xHistidine tag ($H_6$) on the β subunits and a specific cysteine substitution in the γ subunit (γK108C). The soluble $F_1$ portion ($F_1$-ATPase, $F_1$) was isolated from membranes and depleted of subunits δ and ε. $H_6$-tagged ε subunit was overexpressed separately, and contained a unique cysteine mutation, εR99C. Details of the plasmid constructs used, protein expression and purification procedures are given below.

*Plasmid constructs*. Plasmid pJW1 [50] was the basis for the construct used to overexpress all $F_OF_1$ subunits. The $H_6$-tag at the N-terminus of β was engineered previously[51]. All native Cys of β and γ subunits were replaced by Ala by replacing appropriate restriction fragments of pJW1 with the corresponding regions of a similar plasmid encoding Cys-free $F_OF_1$ [52]. The γK108C mutation (AAA > TGC) was made by site-directed mutagenesis (QuikChange II, Agilent Technologies). Wild-type ε has no Cys, and plasmid $pH_6ε$ [53] was mutated to express $H_6$-tagged ε with εR99C (CGT > TGT). The plasmid $pH_6ε$ comprises the amino terminal (His)6 and a rTEV protease site sequence, MSYYHHHHHH-DYDIPTTENLYFQGA, preceding the *atpC* open reading frame[53]. For all tag sites and mutated regions, DNA sequencing was done to confirm the presence of the expected tag/mutant sequence and the absence of any undesired sequence changes.

*Expression of $F_OF_1$ and purification of $F_1(H_6β/γK108C)$ depleted of δ and ε subunits.* For overexpression of the engineered ATP synthase, pJW1($H_6β/γK108C$) was transformed into an *E. coli* strain that lacks a chromosomal *atp* operon (LE392Δ(*atpI-C*) [54]). Cells were grown in up to 10 L of defined medium[50] supplemented with 50 mg Met/L, and everted membranes were prepared[50]. Soluble $F_1(H_6β/γK108C)$ was released from membranes and purified as described[28]. The ε subunit was depleted from $F_1$ with an anti-ε immunoaffinity column[55] as described[56]. To remove residual impurities and deplete most δ subunit, $F_1$ was subjected to immobilized metal affinity chromatography at 4°C. $F_1(H_6β/γK108C)$ was diluted in Talon buffer (50 mM Tris-HCl, 40 mM 6-amino n-hexanoic acid, 10%(v/v) glycerol, 5 mM β-mercaptoethanol [βME], 1 mM ATP; pH adjusted to 7.2 at 22°C) and bound to a 10 ml column of Talon resin (Clontech). The column was washed at 2 ml/min with Talon buffer and different additions as follows: 2 volumes of buffer alone, 4 volumes plus 0.2%(w/v) LDAO to deplete most of subunit δ, 2 volumes of buffer alone to remove LDAO, and 3 volumes plus 0.1 M imidazole to elute $F_1(H_6β/γK108C, –δε)$. The final $F_1$ sample was dialyzed against Talon buffer (with 1 mM DTT instead of βME) to remove imidazole and residual LDAO, then concentrated to >5 mg/ml by ultrafiltration (Vivaspin, 50 kDa MWC) and stored at -80°C.

*Expression and purification of $H_6$-tagged εR99C.* $His_6$-tagged ε subunit ($H_6εR99C$) was expressed in *E. coli* strain BL21Star(DE3) (Life Technologies). Cell growth and induction by IPTG were done as described[53]. The $H_6εR99C$ subunit was purified essentially as described for $H_6ε$[56] but buffers for Talon chromatography contained 1 mM βME, whereas buffers for gel filtration and storage contained 1 mM DTT.

*Labeling of $F_1(H_6β/γK108C)$ with Atto488-maleimide and $H_6εR99C$ with Atto647N-maleimide.* The γ subunit of $F_1$ with the mutation K108C was labeled with Atto488-maleimide as described[57, 58]. Fluorescence labeling of 13 μM $F_1$ resulted in a labeling ratio of 0.55 for the γ subunit according to quantitative SDS-PAGE analysis using a Typhoon scanner. In addition, cysteines of the residual δ subunit were partly labeled with Atto488. $F_1$-Atto488 was flash-frozen in liquid $N_2$ and stored at -80° C in MTKE7 buffer (20 mM MOPS-Tris, 50 mM KCl, 0.1 mM EDTA, pH 7.0) with 10% glycerol and 1 mM ATP. $His_6$-tagged ε subunit with the mutation R99C was labeled with Atto647N-maleimide. 50.1 μM ε contained 17.4 μM Atto647N yielding a labeling ratio of 0.30. ε-Atto647N was flash-frozen in liquid $N_2$ and stored in MTKE7 buffer with 10% glycerol at -80° C.

Single-molecule FRET experiments in the ABEL Trap were carried out in liposome buffer (20 mM succinic acid, 20 mM tricine, 80 mM NaCl, 0.6 mM KCl, 2.5 mM $MgCl_2$, adjusted pH to 8.0 with NaOH) which did not cause sticking of the $F_1$ protein, and caused only minor sticking of the ε subunit alone. Protein aliquots were used as soon as possible, within 12 h after thawing.

## 2.2 ABEL trap for confocal single-molecule FRET measurements in solution

We achieved prolonged observation times of single FRET-labeled $F_1$-ATPase in solution using the ABEL trap as follows.

*ABEL trapping of $F_1$-γ-ε.* ABEL trapping used the knight's tour beam scanning[40] implemented on a FPGA with continuous-wave (CW) excitation provided by the 488 nm line of an argon ion laser (Innova 90, Coherent). A standard 488 dichroic (Semrock) was used to separate excitation and fluorescence. Standard relay optics image the micron-sized grid pattern to the sample plane of an inverted microscope (IX-71, Olympus) equipped with an oil-immersion objective (1.35 NA, 60x, Olympus). Excitation irradiance was 1.2 kW/cm$^2$. FRET was detected by splitting the channels in a typical scheme to separate Atto488 (donor) and Atto647N (acceptor) fluorescence with a 580 dichroic (Semrock). A 150-μm pinhole was added to the detection path, as was a 488-nm notch filter and a 785SP filter before the dichroic (Chroma). Donor channel long pass detection filters included a HQ500LP, and acceptor filters included an HQ630LP (Chroma). Si APDs were used for donor (SPCM-AQ4C, Perkin-Elmer) and acceptor (SPCM-AQR-15, Perkin-Elmer) detection. Arrival times of detected photons were recorded using the FPGA acquisition capability and with TCSPC electronics (PicoHarp 300, Picoquant) in parallel. However, the picosecond time resolution of the PicoHarp 300 was not required for CW measurements of binned fluorescence intensity only.

*$F_1$-γ-ε controls.* Pumping with 638-nm CW excitation was used to perform fluorescence correlation spectroscopy (FCS) on the ε-Atto647N sample alone. This showed rare large aggregates in the sample which adhered to the glass surface, but overall the extracted diffusion time showed that the vast majority of labeled ε were of the ~5 nm hydrodynamic size. Pumping with 488-nm CW excitation for FCS was also performed on the donor-only $F_1$-Atto488 sample. No acceptor channel fluorescence events were observed, indicating that the signal attributed to FRET in the ABEL trap experiments could not arise from impurities or leakage of Atto488 fluorescence into the acceptor channel. These controls confirmed that labeled ε was of the correct size and that FRET was observed in the subsequent ABEL trap experiments.

*Voltage feedback.* ABEL trap microfluidic cells were made by bonding glass coverslips and structured poly(dimethylsiloxane) (PDMS) microfluidic pieces as described previously[38]. Feedback voltages were calculated using custom-written software on an FPGA platform (NI 7833-R, National Instruments) using the knight's tour algorithm with Kalman filtering. Further details have been previously published[40]. Following the feedback calculation, the bipolar voltage outputs from the FPGA were amplified 8x with two high-voltage amplifiers (PS732-K, OK Electronics) before being applied to four platinum electrodes suspended from the microscope and inserted in the solution in the appropriate trap reservoirs.

*Data analysis.* Data analysis was performed in Matlab (The Mathworks, Inc.). Time-tagged detected photons were binned into 10 ms bins to create time traces of emission intensity as shown. When necessary, shifts in intensity and the intervening periods of relatively constant intensity were identified by a change-point algorithm[59]. Because trapping of $F_1$ε was based on fluorescence from the acceptor channel, the change-point algorithm in this case was applied only to the acceptor channel to avoid the intensity spikes in the donor channel arising from donor-only species.

# 3 RESULTS

To monitor the C-terminal conformation of the ε subunit in single $F_1$-ATPase in real time we constructed two cysteine mutants for specific internal distance measurements based on Förster resonance energy transfer (FRET). The $F_1$ part of the holoenzyme $F_oF_1$-ATP synthase from *E. coli* was purified as described previously[28] but with the help of an introduced His-tag at the N-terminus of the β subunits. The γK108C mutation was introduced in a portion of the γ subunit that was expected to be accessible for fast labeling with Atto488-maleimide. In this $F_1$ mutant, native cysteines in the β and γ subunits had been replaced by alanines. During $F_1$ preparation steps, most of the δ and ε subunits were depleted, so that specific labeling of the γK108C cysteine was expected. Atto488-maleimide labeling of F1-γ108C was determined by quantitative SDS-PAGE analysis (Fig. 2). Accordingly, a labeling efficiency of 55% for the γ subunit was achieved. Unbound dye was removed by centrifuge columns. However, the remaining substoichiometric amount of the δ subunit was partly labeled *via* its native cysteines, shown as a faint fluorescent band with lower molecular weight in lane 2 of Fig. 2. Comparing the fluorescence intensities of labeled γ and δ subunits in the gel revealed that approximately 15% of $F_1$ was labeled at the δ subunit. ATP hydrolysis activity of labeled $F_1$-ATPase was measured at

30° C to reveal functionality. A photometric assay was used to determine the concentration of the released phosphate. We found an activity for the $F_1$-ATPase of about 260 s$^{-1}$, which is in the same range as published activities for unlabeled $F_1$[56, 60].

Subunit ε99C was constructed with an N-terminal peptide extension comprising a His-tag for protein purification and a TEV protease cleavage site for optional post-processing. His-tags and protease cleavage site increased the molecular weight from 14.9 kD for the native ε subunit to 18.1 kDa. Purified subunit ε was labeled with Atto647N-maleimide that added about 770 Da to the molecular weight of ε and one additional positive charge upon reaction with the cysteine (according to the data table of the supplier Atto-tec, Germany). Atto647N labeling efficiency was determined to 30%, and unbound dye was removed by dialysis. SDS-PAGE revealed a main fraction of Atto647N-labeled ε with an apparent molecular weight in the range of 20 kDa (indicated as ε* in lane 3 of Fig. 2), and a second labeled protein band at around 16 kDa (ε** in lane 3 of Fig. 2). This sample was used in the subsequent ε binding to $F_1$-γ-atto488 without separating of ε* from ε**.

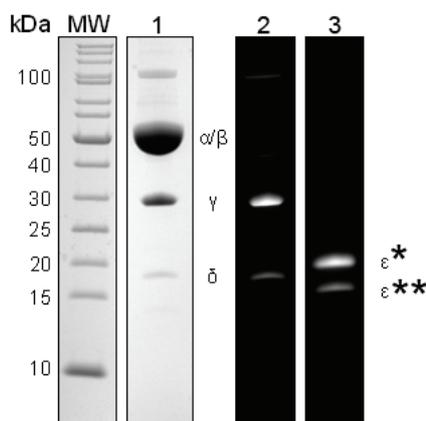

**Figure 2.** SDS-PAGE of labeled $F_1$-γ-atto488 and ε99-atto647N (12% PA gel according to Schaegger and von Jagow[61]), with Coomassie Blue staining of $F_1$-ATPase subunits in lane 1, and fluorescence images in lanes 2 and 3. Lanes 1 and 2 show $F_1$-γ108-atto488. Atto488-maleimide labeling also occurred on cysteines of the residual fraction of δ subunit. Lane 3, ε99-atto647N preparation showing two fluorescence-labeled products ε* and ε** (for details see text). Molecular weight standard on the left.

Due to ε's high affinity ($K_D$~ 0.3 nM[56]), mixing 3 μM $F_1$-γ-atto488 with ε-atto647N (4 μM) yielded quantitatively the recombined $F_1$-γ-ε. Given the fluorophore labeling efficiencies, only 30% of the FRET donor-labeled $F_1$-γ-atto488 could bind a FRET acceptor-labeled ε-atto647N. The maximum observable mean FRET efficiency is decreased further by the contribution of 15% Atto488 fluorophores bound to δ. These Atto488 dyes will not contribute to FRET in $F_1$-γ-ε because the cysteines of the δ subunit are expected to be more than 8 nm away from a FRET acceptor dye on ε99. The Förster radius of the unbound fluorophores Atto488-Atto647N was given as $R_0$ = 5.1 nm according to the supplier Atto-tec.

Cysteine positions in γ and ε of $F_1$ had been designed for FRET fluorophore distances including the linkers to the cysteine residues of approximately 3 nm in the case of a C-terminal conformation of ε representing the 'extended up' structure[28] (see Fig. 1 B), or ~6 nm in the case of the 'down' conformation of ε (see Fig. 1 C). The 'extended up' structure of ε inhibits ATP hydrolysis of $F_1$-ATPase by preventing γ subunit rotation. In contrast, the active $F_1$ enzyme with rotating γ and ε subunits during ATP turnover is expected to have the C-terminal helices of ε in a 'down' conformation. A conformational transition from the inhibited to the active $F_1$-ATPase will be correlated with a FRET efficiency change.

Therefore we first measured fluorescence spectra of the reassembled $F_1$-γ-atto488/ε-atto647N in cuvettes. Using excitation at 488 nm, the emission spectra revealed FRET for various biochemical conditions. Examples are shown in

Fig. 3. In the absence of AMPPNP, FRET between γ-atto488 and ε-atto647N within $F_1$ was indicated by the fluorescence peak with a maximum at 661 nm. The relative intensity of 0.16 for this FRET-sensitized Atto647N emission maximum was compared to the normalized fluorescence maximum of the FRET donor Atto488 at 520 nm. Addition of 1 mM $Mg^{2+}$AMPPNP, a non-hydrolyzable ATP analog that binds to $F_1$-ATPase, caused a time-dependent decrease of Atto647N emission to 0.135 in the first minutes after mixing (light grey curve in Fig. 3), and to 0.105 after 15 minutes (dark grey curve in Fig. 3) and 30 minutes (data not shown). Calculating normalized spectra in Fig. 3 compensated the effects of dilution and of fluorescence loss due to sticking of $F_1$ to the cuvette surfaces.

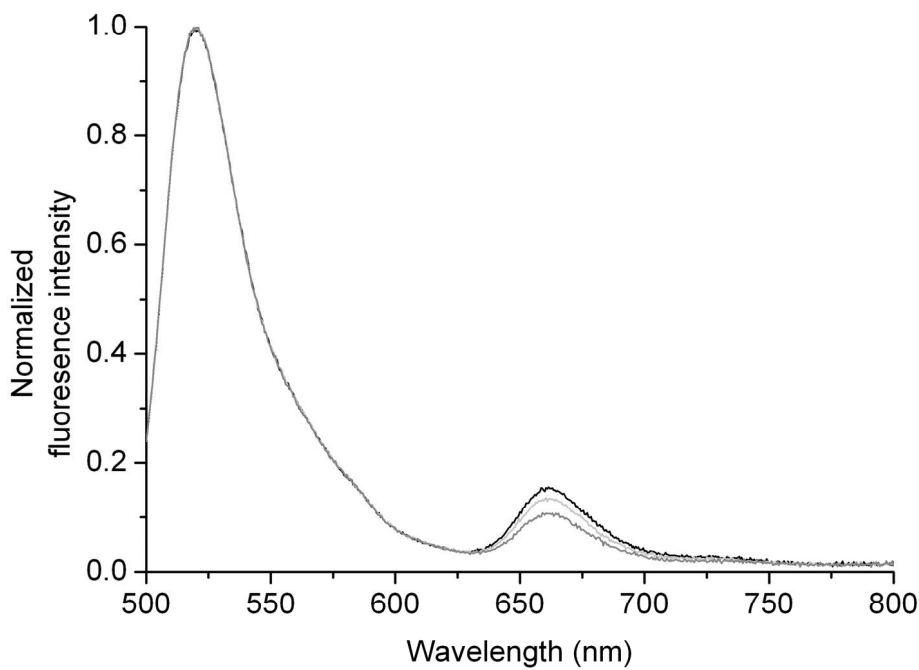

**Figure 3.** Time-dependent normalized fluorescence spectra of labeled $F_1$-γ-atto488/ε99-atto647N in the absence and presence of 1 mM $Mg^{2+}$ AMPPNP. Protein concentrations were 15 nM for $F_1$ and 20 nM for ε. Excitation was achieved with 488 nm using 5 nm slits at excitation and emission monochromators. Black line, fluorescence spectrum in the absence of AMPPNP; light grey, fluorescence spectrum in the presence of AMPPNP immediately after nucleotide addition and mixing ("0" min); dark grey, fluorescence spectrum in the presence of AMPPNP 15 min after nucleotide addition.

Smaller decreases of the relative intensity of FRET acceptor Atto647N emission were observed also upon addition of 1 mM $Mg^{2+}$ATP or 1 mM $Mg^{2+}$ADP. An apparent equilibrium for FRET changes was reached after 20 to 30 minutes. The detergent LDAO disrupts inhibition by ε's CTD[62]. Adding 0.01% LDAO to 1 mM $Mg^{2+}$ATP caused a further decrease of FRET acceptor intensity. In the presence of 0.1% LDAO and 1 mM $Mg^{2+}$ATP, FRET acceptor emission was almost completely lost (not shown). Changes in the relative FRET acceptor fluorescence could be interpreted as a conformational change of the C-terminus of subunit ε, that is, from a shorter distance between Atto488 on γ108 to Atto647N on ε99 to a larger distance for these dyes. However, partial dissociation of the ε subunit from $F_1$ in the presence of nucleotides could also result in the loss of FRET acceptor emission.

Measuring distances and distance changes within single molecules circumvents this ambiguity of ensemble averaging. In a preliminary experiment[57], we diluted $F_1$-γ-atto488/ε-atto647N to less than 1 nM immediately before starting single-molecule FRET measurements in solution on a confocal microscope. Diffusion of $F_1$ (~10 nm diameter) was fast, i.e. about 3 ms on average through an enlarged confocal detection volume with a size of a few femtoliters (characterized by a diffusion time of ~300 μs for a free fluorophore in buffer). Therefore, the short observation times only allowed us to determine an average FRET distance for each enzyme, but not time-dependent distance changes.

Subsequently we investigated the conformations of the C-terminal helices of ε in $F_1$ held in solution by the ABEL trap. Because the ε subunit was found to dissociate from $F_1$ alone with dissociation constant in the ~nM range[56], a specialized scheme was required for ABEL trapping. The ~pM concentrations required for trapping $F_1$-γ-Atto488/ε-Atto647N meant that significant excess concentrations of either $F_1$-γ-Atto488 and/or ε-Atto647N were required under trapping conditions. Therefore, we chose a scheme in which both F1-γ-Atto488 and ε-Atto647N were present in the ABEL trap at higher concentrations, i.e. in the range of 1-10 nM. We then excited the FRET donor $F_1$-γ-Atto488 continuously with 488 nm (CW excitation), but trapping was based only on the FRET acceptor fluorescence arising from ε-Atto647N. This allowed selective trapping of only the $F_1$-γ-Atto488/ε-Atto647N complex, at the price of many short 'spikes' from $F_1$-γ-Atto488 diffusing through the trapping scan pattern without being trapped.

Figure 4 shows several trapped $F_1$ enzymes each in the absence and presence of ATP, respectively. Donor (dark grey) and acceptor (light grey) intensities are shown with the FRET efficiency E (the proximity factor), without corrections for detection efficiencies of the setup and quantum yields of the fluorophores (black dotted lines). Most trapped $F_1$-γ-Atto488/ε-Atto647N enzymes showed a single FRET level (Fig. 4 a, g, h, i), but transitions to higher or lower FRET levels were also observed (Fig. 4 c, d, e, k, l). The transitions, however, were too rare to build up meaningful statistics. Large aggregates of $F_1$ were also observed (Fig. 4 f, m). No obvious difference was observed in the conformational dynamics upon addition of 1 mM ATP to initiate ATPase turnover, but this is not surprising since ε inhibition is noncompetitive *vs.* ATP (see discussion). Control experiments on the donor-only $F_1$-γ-Atto488 in the ABEL trap showed that the donor-channel intensity of $F_1$-γ-Atto488 with no FRET occurring was 11.5 kHz, meaning that total count rates greater than ~20 kHz were $F_1$ aggregates or $F_1$ that also had δ-Atto488 labeling. These bright photon bursts were excluded from further processing. In general, the excitation irradiance used for these experiments was higher than in other single-dye experiments in the ABEL trap because we were mostly interested in the overall FRET level for each $F_1$ with the idea of building up the FRET distribution, for which long time observation was less critical. Preliminary experiments at lower irradiance in quartz cells (for low background) yielded similar overall FRET dynamical behavior when $F_1$ was trapped for longer ~s times, but again with no obvious changes upon addition of ATP.

Histograms of the FRET levels of the identified change-point levels in three different conditions are shown in Figure 5. The conditions tested were the presence and absence of ATP, AMPPNP or ADP. The total number of FRET levels contributing to the normalized histogram in Fig. 5 was 2697 in the absence of added nucleotides, 759 in the presence of ATP and 560 in the presence of AMPPNP. In the absence of added nucleotides, the main fraction of $F_1$-γ-Atto488/ε-Atto647N exhibited a proximity factor of 0.6. Upon adding ATP no significant difference in the distribution of FRET states could be detected, with most of the FRET-labeled $F_1$ enzymes showing a proximity factor of 0.6. In the presence of AMPPNP, fewer change-point levels contributed to the histogram that comprised an increased fraction of $F_1$ with lower proximity factors (0.2 to 0.3) as well as higher FRET (~0.9), at the expense of a decreasing fraction of $F_1$ with proximity factor of 0.6. In the presence of ADP, the FRET level distribution (data not shown) appeared similar to the AMPPNP distribution. The increased width in the ADP and AMPPNP histograms is not necessarily due to conformational dynamics only, but may be due to the increased impurity concentration as observed under these measurement conditions. The high excess concentration of $F_1$-γ-Atto488 meant that fluctuations in the donor channel were substantial due to diffusing $F_1$-γ-Atto488 that were not trapped. Because the time average of these intensity 'spikes' over the ~100 ms trapping duration was a constant offset, we expected it not to affect the means of the distributions or dynamics of the observed FRET states. Some of the broadness in the histograms, however, almost certainly resulted from the large intensity spikes in the donor channel.

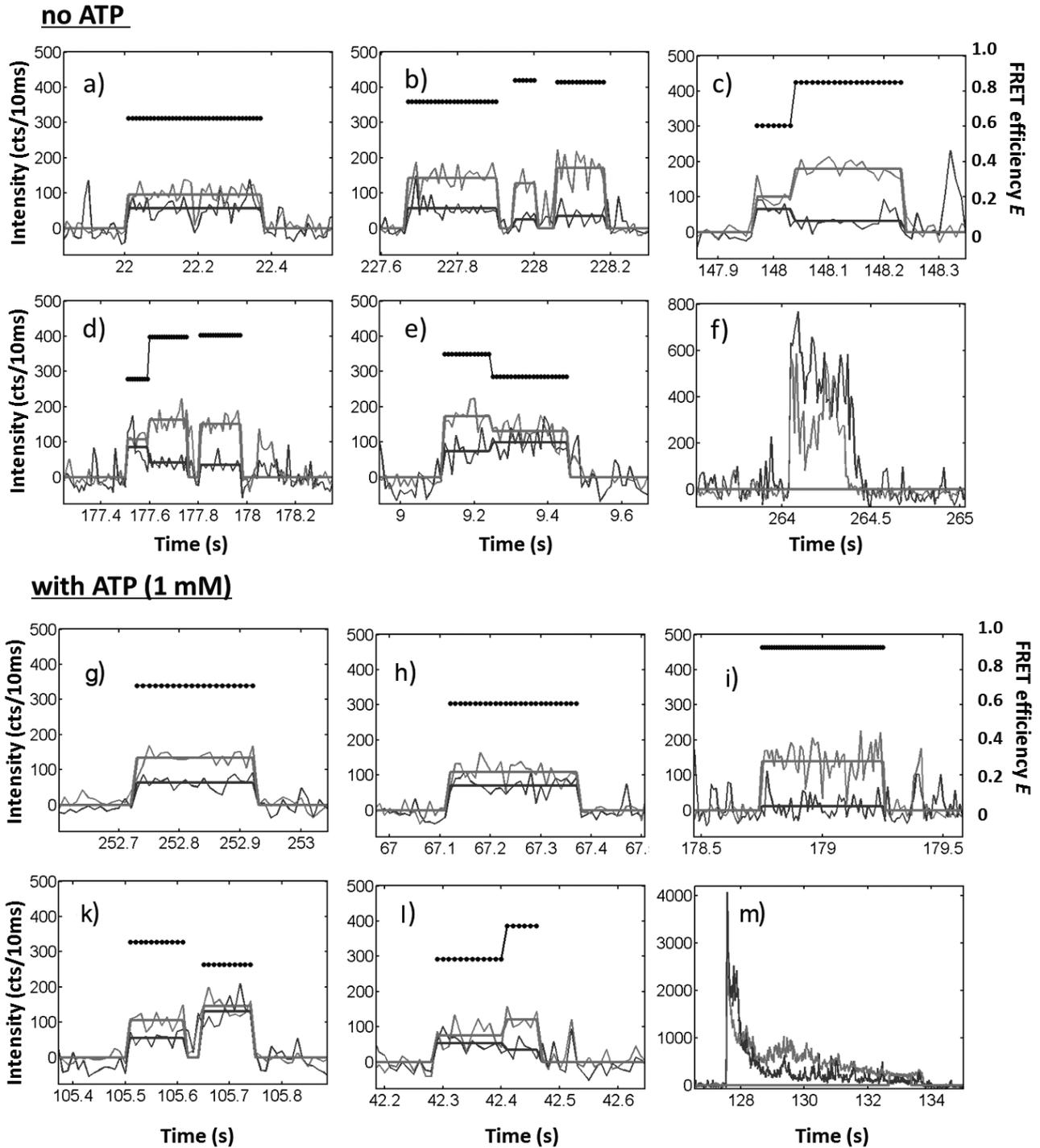

**Figure 4.** Representative ABEL trapped $F_1$-$\gamma$-Atto488/$\varepsilon$-Atto647N. Intensity in the donor (dark grey) and acceptor (light grey) channels is plotted, together with change-point levels overlaid (heavy lines) and FRET efficiency $E$, calculated as the proximity factor (dotted lines). FRET was clearly observed in the trapped proteins shown (a - f) in the absence of nucleotides and (g - m) in the presence of 1 mM $Mg^{2+}$ATP. Most proteins exhibited only one FRET state (a, g, h, i). Rarely, trapped proteins showed transitions to higher or lower FRET (c, d, e, k, l). Large aggregates were often seen (f, m) and were excluded from any further analysis.

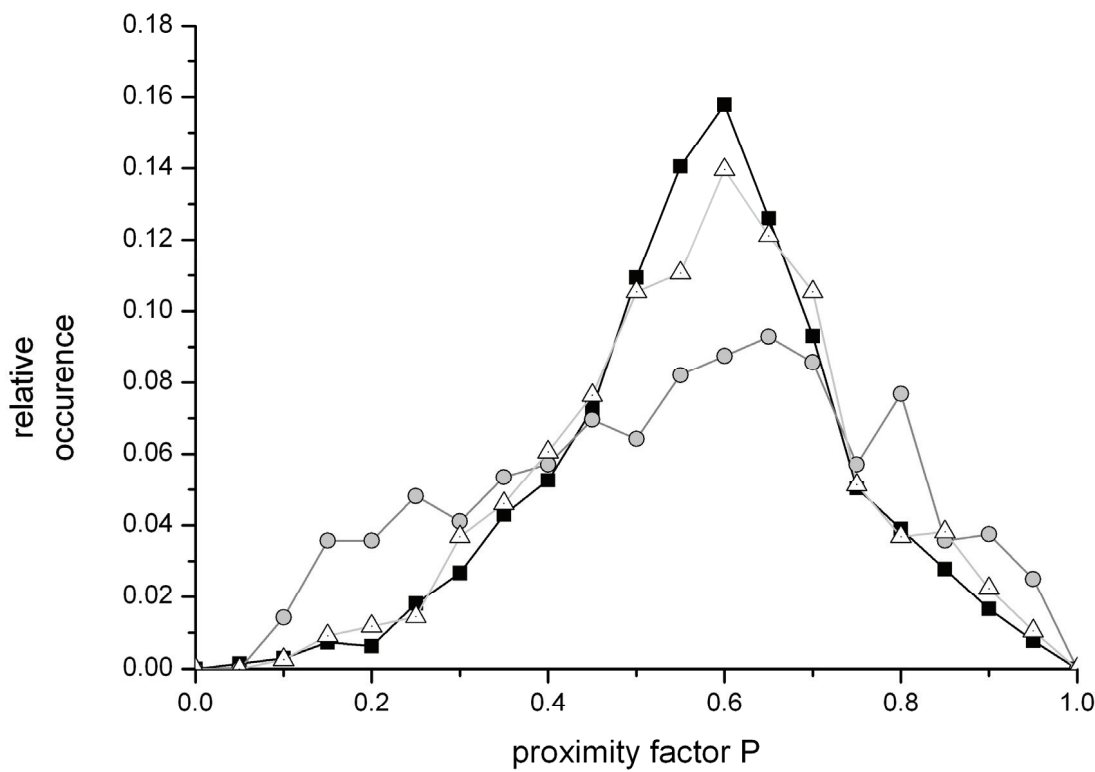

**Figure 5.** Normalized proximity factor histograms based on change-point levels for ABEL-trapped $F_1$-$\gamma$-Atto488/$\varepsilon$-Atto647N under various conditions. No difference was observed in the absence (black squares) and presence of 1 mM $Mg^{2+}$ATP (white triangles). Addition of 1 mM $Mg^{2+}$AMPPNP resulted in a relative shift of the FRET level population with proximity factors P ~0.6 towards low FRET efficiencies (P ~0.2 - 0.3) as well as very high proximity factors P ~0.9 (dark grey dots).

## 4 DISCUSSION

Conformational changes of soluble proteins like $F_1$-ATPase can be monitored using two specifically attached fluorophores for an internal distance ruler based on FRET. Because surface attachment might perturb the conformational dynamics, single-molecule FRET data are measured in solution. However, the observation time of a 10-nm sized protein in a confocal detection volume is less than 5 ms due to Brownian motion and, accordingly, long time trajectories with several FRET transitions for an individual protein cannot be recorded in solution.

Instead, conformational dynamics of soluble proteins can be recorded as single snapshots of one conformation. In a previous preliminary study[57] different FRET levels were found for $F_1$-$\gamma$-Atto488/$\varepsilon$-Atto647N enzymes freely diffusing in buffer solution. The observation time of the diffusing $F_1$-ATPase was approximately in the range of 1 to 3 ms using a confocal detection volume with a size of a few femtoliters. Addition of different nucleotides in the presence of $Mg^{2+}$ resulted in distinct FRET efficiency ($E_{FRET}$) distributions. Biochemical data showed that MgADP and $P_i$ stabilize the $\varepsilon$-inhibited state[56]. Addition of ADP/$P_i$ resulted in a dominant population with $E_{FRET}$ about 0.6, similar to the conditions without adding nucleotides. This should represent the $\varepsilon$-inhibited, 'up' state, as in the *E. coli* $F_1$ structure[28]. Adding AMPPNP resulted in an additional population of $E_{FRET}$ about 0.25. This lower $E_{FRET}$ value, therefore, should be the

'down' conformation of the C-terminal domain of ε. However, the majority of $F_1$ complexes were still found at $E_{FRET}$ ~0.6. This was interpreted as a relation to the strong inhibition of isolated $F_1$ by ε[56].

We are interested in long time trajectories of FRET changes of ε in $F_1$-ATPase for a future comparison with the liposome-reconstituted holoenzyme $F_oF_1$-ATP synthase from *E. coli*. Measured with one active enzyme at a time, FRET trajectories obtained in an ABEL trap could provide the required statistical basis for a quantitative description of the rates for the conformational dynamics of the C-terminus of ε. Our FRET-labeled $F_1$-γ-atto488/ε-atto647N was assembled using $F_1$ with cysteine mutation γK108C and depleted of δ and ε subunits, and an engineered ε subunit comprising a cysteine mutation εR99C and extended by His-tags and a protease cleavage site at the N-terminus.

Labeling specificity for the FRET donor Atto488 at the γ subunit (about 55%) was reduced by additional labeling of a remaining fraction of the δ subunit. About 15% of $F_1$ contained Atto488-labeled δ, resulting in a fraction of double-labeled $F_1$ with two Atto488 fluorophores. Thus, $F_1$-ATPase could either contain one or two FRET donor dyes on γ and on δ. However, labeled δ was expected to yield either no FRET or only a very low FRET efficiency in the presence of ε99-atto647N because of the large distance of 8 nm or more. Specificity of labeling for the full-length ε was compromised by a fraction of ε with lower molecular weight as shown in Fig. 2. We could not determine how this small fraction (about 20%) of ε was truncated during protein preparation and where the cleavage might have occurred. The FRET-labeled $F_1$ could be assembled with two different types of ε which could in principle affect the FRET efficiency as well.

Given the 30% labeling efficiency of ε, a contribution of labeled δ to the donor-only fraction of $F_1$, fluorescence quantum yields of 0.80 for Atto488 and 0.65 for Atto647N (according to Atto-tec), we can estimate the relative intensity of the FRET-sensitized Atto647N emission in a bulk fluorescence spectrum for a fluorophore distance in the range of the Förster radius $R_0 = 5.1$ nm. In the normalized emission spectrum, these simplified assumptions would result in a relative intensity of the FRET acceptor maximum by multiplication of all factors 0.3*0.8*(0.65/0.80) = 0.195. In the measured emission spectra in Fig. 3, a relative fluorescence intensity of 0.16 for FRET-sensitized Atto647N was found. Repeating the FRET spectroscopy measurements showed similar values for FRET-sensitized emission of Atto647N indicating a quantitative and stable assembly of $F_1$-γ-atto488 with ε-atto647N from micromolar concentrations of mixing down to 15 nM ($F_1$) and 20 nM (ε) during fluorescence measurements. However, the total fluorescence intensity decreased over time due to sticking of $F_1$ and/or ε to the glass surfaces of the cuvettes as known from previous spectroscopy and FCS experiments[63].

Adding nucleotide (especially AMPPNP) to the $F_1$-γ-Atto488/ε-Atto647N enzymes in the cuvette resulted in a relative loss of FRET acceptor intensity. This could be interpreted as an induced conformational change of the C-terminus of ε accompanied with a mean increase in dye distances or lowered mean FRET efficiency, respectively. However, an alternative explanation could be the nucloetide-dependent dissociation of ε from $F_1$.

A recent study monitored bulk kinetics of *E. coli* $F_1$/ε binding and dissociation as a means to probe the ability of $F_1$-bound ε to switch its CTD to and from a tightly-bound inhibitory conformation, and to test the effects of different catalytic-site ligands on ε's conformational changes[56]. Inhibition of isolated *E. coli* $F_1$-ATPase by ε is noncompetitive *vs* ATP substrate, and that study confirmed that turnover of ATP hydrolysis has little effect on ε's strong bias towards its inhibitory state (*i.e.*, $K_I$ is ~0.6 nM for ε inhibition of ATPase, and $K_D$ is ~0.3 nM for $F_1$/ε binding in the absence of $Mg^{2+}$ATP). It also confirmed that the ε CTD switches to the inhibitory state at the post-hydrolysis catalytic dwell and that hydrolysis products, $Mg^{2+}$ADP and Pi, stabilize the high-affinity inhibitory state of ε. In contrast, addition of the nonhydrolyzable analog AMPPNP rapidly (<10 s) induced $F_1$/ε complexes to dissociate ~80-fold faster, comparable to dissociation when the ε CTD was absent. However, the bulk study could not discriminate whether AMPPNP binding (i) directly altered the kinetics of ε's switch to or from the inhibitory conformation or (ii) induced $F_1$ complexes in the active state to undergo a conformational or rotational change that prevented the ε CTD from returning to its inhibitory conformation. Such questions could eventually be answered by sm-FRET studies like those being developed in the present study.

Accordingly, concentrations of $F_1$ and ε in the range of 10 nM, i.e. above the dissociation constant of ε, were required to measure single FRET-labeled $F_1$-γ-Atto488/ε-Atto647N in the ABEL trap. Trapping FRET-labeled $F_1$-ATPase was achieved with the feedback based on the fluorescence (photons) of the FRET acceptor Atto647N. Due to the high $F_1$ concentration in the trapping region, a fluctuating background from donor-only labeled $F_1$-$\gamma$-Atto488 was always present. To identify photon bursts and to assign FRET levels without model assumption, a change-point algorithm was applied[59] to the time trajectories with 10 ms binning. Background was subtracted using averaged donor and acceptor intensities before and after the ABEL-trapped $F_1$-ATPase. Within the photon burst of a $F_1$-ATPase, the change-point algorithm searched for areas with constant intensities to assign FRET levels without models. Photon bursts with constant FRET efficiencies as well as fluctuating FRET levels were found for all biochemical conditions. All FRET levels were summarized in proximity factor histograms for the various nucleotide conditions.

The proximity factor histograms in the absence and presence of 1 mM $Mg^{2+}$ATP were similar. The dominant proximity factor of ~0.6 corresponds to significant FRET between residues $\gamma$108 and $\varepsilon$99, which we interpret to be the $\varepsilon$ 'up' state. Therefore, this CTD conformation of $\varepsilon$ in the 'up' state, i.e. preventing $\gamma$ subunit rotation, is associated with inhibiting ATP hydrolysis of $F_1$-ATPase in solution. Fluctuating FRET levels indicated that the enzyme is only transiently activated in the presence of $Mg^{2+}$ATP. These ABEL-trap results are consistent with bulk findings that conditions for ATP hydrolysis do not shift $\varepsilon$ from the inhibited, tightly-bound state. In a preliminary smFRET study[57], the more significant fraction of low-FRET $F_1$-$\gamma$-Atto488/$\varepsilon$-Atto647N may have arisen slowly, as some post-hydrolysis complexes eventually convert to the $Mg^{2+}$ADP-inhibited state of the enzyme, which competes with the inhibitory switch of $\varepsilon$'s CTD.

Upon addition of $Mg^{2+}$AMPPNP, a small fraction of $F_1$-ATPase in the ABEL trap showed lower proximity factors around 0.3 at the expense of the 0.6 fraction. This low-FRET population corresponds to larger distances between $\gamma$108 and $\varepsilon$99, i.e. to the CTD 'down' conformation of $\varepsilon$. However, the relative occurrence of this conformation is less than might be expected from the bulk effects of AMPPNP on $F_1$/$\varepsilon$ dissociation[56]. Likewise, AMPPNP did not reduce the mean FRET efficiency (Fig. 3) as much as expected, especially considering that the bulk change in F1/$\varepsilon$ off-rate should have shifted the composition from ~96% bound F1/$\varepsilon$ to only ~50% ($K_D$ shift from 0.3 to ~17 nM). However, the experiments on bulk $F_1$/$\varepsilon$ dissociation[56] were done in conditions to minimize possible rebinding of $F_1$/$\varepsilon$, with no free $F_1$ or $\varepsilon$ in the dissociation step, whereas the current ABEL trapping was done with significant nM concentrations of $F_1$ and $\varepsilon$ so that rebinding could occur, and $\varepsilon$ might have had a chance to re-form the inhibitory state even with AMPPNP present.

It is useful to consider what improvements in the experimental design might yield a more convincing observation of $\varepsilon$ inhibition. Given the nucleotide-dependent binding constants of $\varepsilon$ to $F_1$, even possible improvements for ABEL trapping FRET-labeled $F_1$-ATPase such as increased labeling efficiencies on $\gamma$ and $\varepsilon$, a complete removal of $\delta$, separation of the different $\varepsilon$ subunit fragments (or aggregates), longer observation times by reducing photobleaching of the dyes, and the use of quartz microfluidics may not fully overcome these limitations. Instead, the holoenzyme $F_o F_1$-ATP synthase reconstituted in liposomes could be studied in the ABEL trap. In previous single-molecule FRET studies on subunit rotation in $F_o F_1$-ATP synthase, the labeled $\varepsilon$ subunit did not dissociate at concentrations of less than 100 pM[21, 23, 27, 64, 65]. The reconstituted holoenzyme will also allow variation of the pH difference and electric potential across the membrane, which are required for ATP synthesis. Changes in pH difference and electric potential could mimic the critical cellular conditions for regulating $\varepsilon$'s C-terminal domain conformation to prevent wasteful ATP hydrolysis. Monitoring the conformational dynamics using single-molecule FRET and for extended periods of time should help to unravel this essential control mechanism of the enzyme.

**Acknowledgements**

This work was supported in part by NIH grant R01GM083088 to T. M. D. and D.O.E grant DE-FG02-07ER15892 to W. E. M. Additional financial support from DFG grant BO 1891/16-1 to M. B. is gratefully acknowledged. We thank Marcus L. Hutcheon (now at Bristol-Myers Squibb, Syracuse, NY) for purification of proteins used in this study.